\documentclass[journal]{IEEEtran}
\usepackage{amsmath}
\usepackage{amssymb}
\usepackage{graphicx}
\usepackage{color}
\usepackage{algorithm}
\usepackage{algpseudocode}
\usepackage{array}
\usepackage{booktabs}
\usepackage{tabularx}
\usepackage{textcomp}

\ifCLASSINFOpdf
\else
\fi
\begin{document}
%
% paper title
% Titles are generally capitalized except for words such as a, an, and, as,
% at, but, by, for, in, nor, of, on, or, the, to and up, which are usually
% not capitalized unless they are the first or last word of the title.
% Linebreaks \\ can be used within to get better formatting as desired.
% Do not put math or special symbols in the title.
\title{Bare Demo of IEEEtran.cls\\ for IEEE Journals}
%
%

% note the % following the last \IEEEmembership and also \thanks - 
% these prevent an unwanted space from occurring between the last author name
% and the end of the author line. i.e., if you had this:
% 
% \author{....lastname \thanks{...} \thanks{...} }
%                     ^------------^------------^----Do not want these spaces!
%

% The paper headers
\markboth{Journal of \LaTeX\ Class Files,~Vol.~14, No.~8, August~2015}%
{Shell \MakeLowercase{\textit{et al.}}: Bare Demo of IEEEtran.cls for IEEE Journals}
% The only time the second header will appear is for the odd numbered pages
% after the title page when using the twoside option.
% 
% *** Note that you probably will NOT want to include the author's ***
% *** name in the headers of peer review papers.                   ***
% You can use \ifCLASSOPTIONpeerreview for conditional compilation here if
% you desire.

% If you want to put a publisher's ID mark on the page you can do it like
% this:
%\IEEEpubid{0000--0000/00\$00.00~\copyright~2015 IEEE}
% Remember, if you use this you must call \IEEEpubidadjcol in the second
% column for its text to clear the IEEEpubid mark.

% use for special paper notices
%\IEEEspecialpapernotice{(Invited Paper)}

\title{Zero-Knowledge Proof-Based Consensus for Blockchain-Secured Federated Learning}
\author{Tianxing Fu, Jia Hu, Geyong Min, Zi Wang}
% make the title area
\maketitle

% As a general rule, do not put math, special symbols or citations
% in the abstract or keywords.
\begin{abstract}
Federated learning (FL) enables multiple participants to collaboratively train machine learning models while ensuring their data remains private and secure. Blockchain technology further enhances FL by providing stronger security, a transparent audit trail, and protection against data tampering and model manipulation. 
Most blockchain-secured FL systems rely on conventional consensus mechanisms: Proof-of-Work (PoW) is computationally expensive, while Proof-of-Stake (PoS) improves energy efficiency but risks centralization as it inherently favors participants with larger stakes. Recently, learning-based consensus has emerged as an alternative by replacing cryptographic tasks with model training to save energy. However, this approach introduces potential privacy vulnerabilities, as the training process may inadvertently expose sensitive information through gradient sharing and model updates.
 To address these challenges, we propose a novel Zero-Knowledge Proof of Training (ZKPoT) consensus mechanism. This method leverages the zero-knowledge succinct non-interactive argument of knowledge proof (zk-SNARK) protocol to validate participants’ contributions based on their model performance, effectively eliminating the inefficiencies of traditional consensus methods and mitigating the privacy risks posed by learning-based consensus. Based on this consensus mechanism, we design a blockchain-secured FL system, integrating a ZKPoT-customized block and transaction structure alongside IPFS to streamline the FL and consensus processes while significantly reducing communication and storage costs.
 We analyze our system's security, demonstrating its capacity to prevent the disclosure of sensitive information about local models or training data to untrusted parties during the entire FL process. Extensive experiments demonstrate that our system is robust against privacy and Byzantine attacks while maintaining accuracy and utility without trade-offs, scalable across various blockchain settings, and efficient in both computation and communication.
\end{abstract}

\begin{IEEEkeywords}
Federated learning, Blockchain, Zero-knowledge proof, Edge computing, Consensus mechanism
\end{IEEEkeywords}

\IEEEpeerreviewmaketitle
\renewcommand{\thefootnote}{}
\footnote{\textit{Tianxing Fu, Jia Hu, and Geyong Min are with the Department of Computer Science, University of Exeter, Exeter EX4 4QJ, U.K. (e-mail: {tf407; j.hu; g.min}@exeter.ac.uk) Zi Wang is currently with the University of Electronic Science and Technology of China. (e-mail: zi.wang@uestc.edu.cn)}}

\section{Introduction}\label{sec:intro}
\IEEEPARstart{T}{he} conventional deep learning (DL) training process involves centralized collection and processing of vast amounts of local private data, raising significant privacy concerns \cite{al2019privacy}. 
To address this issue, federated learning (FL) \cite{fedavg} was introduced as a promising solution, enabling distributed model training by aggregating locally trained models on a central server without collecting local data.
However, the central server is at risk of single-point failures, making it susceptible to potential collusion or tampering \cite{acm_survey}. On the client side, FL is vulnerable to poisoning and privacy attacks by malicious clients, which jeopardize the system performance and trustworthiness \cite{rao2024privacy}. 
To mitigate these risks, the model aggregation process can be shifted from the central server to a decentralized network of nodes (e.g., edge servers), significantly reducing the likelihood of collusion and tampering. 
Moreover, this decentralized approach can be enhanced in terms of reliability and secured by leveraging the blockchain technology \cite{nguyen2021federated, feng2021blockchain, jin2023lightweight}.
As a decentralized ledger, blockchain ensures data consistency across a network of distributed nodes and offers salient security features including tamper-proof records and traceability, making it difficult for any individual blockchain node to alter data without detection, thereby strengthening defenses against malicious behaviors within the network \cite{acm_survey}. 
Blockchain's security features are underpinned by consensus mechanisms, such as Proof-of-Work (PoW) and Proof-of-Stake (PoS) \cite{Blockchained_on_device}, which enable distributed nodes to reach an agreement on the state of the network without relying on a central authority. PoW requires participating nodes (miners) to solve complex mathematical puzzles to validate transactions and select leaders, while PoS selects validators based on the amount of stake (e.g., cryptocurrency) they hold and are willing to ``stake" as a guarantee that they will act honestly in the network.

Recently, some blockchain-secured FL systems have demonstrated substantial improvements in security and robustness against various attacks \cite{biscotti, commitee_consensus, feng2021bafl, yang2024blockchain, xu2022spdl}.
For example, several studies \cite{Blockchained_on_device, cognitive_com} integrate FL with PoW-based blockchains. 
While enhancing security, PoW consumes substantial computational resources that could otherwise be utilized for FL model training. 
This is particularly problematic for resource-constrained edge devices, as they may struggle to handle the intensive cryptographic puzzles required by PoW.
Additionally, Proof-of-Stake (PoS)-based systems, although more energy efficient, may introduce centralization risks by favoring participants with larger stakes, which could undermine fairness and security \cite{acm_survey}.
Consensus mechanisms based on model performance, such as Proof of Deep Learning (PoDL) \cite{podl} and Proof of Federated Learning (PoFL) \cite{pofl}, offer promising solutions to address this issue. 
These mechanisms replace the computationally intensive mathematical puzzles in PoW with DL training tasks, selecting the client with the best model as the leader. 

However, to ensure the soundness of the leader selection, verifying clients’ model performance is necessary. This verification process typically requires clients to access and run others’ models on a shared test dataset, which inadvertently leads to information leakage, as model parameters might reveal insights derived from the training data \cite{rao2024privacy, DLG}.
To address the above privacy risks, several studies \cite{PoQ, pf-pofl} apply differential privacy (DP) techniques, which introduce Gaussian noise during training, to safeguard model parameters. Nonetheless, existing research \cite{rao2024privacy, wei2020federated, DMIA} indicates that while DP helps in maintaining model confidentiality, it typically results in longer training and aggregation times, as well as lower model accuracy, due to the noise added to the gradients of the target classifier during training. Furthermore, several studies have shown that DP is insufficient in fully preventing privacy attacks. Gupta \textit{et al.} \cite{gupta2022recovering} conducted a gradient inversion attack (GIA) to recover training data from model parameters, while Truex \textit{et al.} \cite{DMIA} performed a membership inference attack (MIA) to determine whether certain data points are within the training dataset, both targeting at models protected by DPs. 

To address these challenges, we propose a novel consensus mechanism, namely Zero-Knowledge Proof of Training (ZKPoT), which enables clients to prove their training results without sharing models. Recently, zero-knowledge proofs (ZKPs) have been used to demonstrate the integrity of inference results of DL models \cite{zkcnn, fan2024validcnn}. As a cryptographic technique, ZKP allows a prover to convince a verifier of the truth of a statement without revealing any information beyond the statement itself.
Specifically, ZKPoT leverages a specific type of ZKP, known as the zero-knowledge succinct non-interactive argument of knowledge (zk-SNARK). Using the zk-SNARK scheme, clients generate cryptographic proofs that encapsulate both the model’s accuracy and the results of the inference computation. These proofs validate the correctness of the inference without revealing the underlying model parameters or sensitive training data, thereby verifying performance while ensuring privacy. Moreover, these proofs are stored on a blockchain, ensuring they remain immutable and can be easily verified by any client within the network. 
To reduce the computational burden for clients, ZKPoT centralizes essential, repetitive proof generation operations required for verifying model updates to a trusted third party - the publisher of FL tasks, which initiates the training process and provides a set of test data \cite{jin2023lightweight}.
ZKPoT also incorporates the Inter-Planetary File System (IPFS) to store large files, such as the global model and proofs, to reduce on-chain communication overhead and storage costs.

The major contributions of this paper are summarized as follows:
\begin{itemize} 
    \item We propose a zero-knowledge Proof of Training (ZKPoT) consensus mechanism for blockchain-secured FL systems. By selecting leaders based on model performance, it significantly reduces the need for extensive computations during consensus. Furthermore, the use of zero-knowledge proofs (ZKPs) allows clients to validate their contributions without exposing their models, enabling them to bypass defense mechanisms like differential privacy (DP) that requires trade-offs in model accuracy.
    \item Based on the ZKPoT consensus, we design a novel blockchain-secured FL system. This system features ZKP-focused block and transaction structures tailored to enhance the efficiency and scalability of the consensus process. It eliminates central server bottlenecks by distributing the model aggregation process across a network of edge nodes. Furthermore, our system substantially reduces on-chain communication and storage costs by leveraging IPFS to store models and proofs and only recording their addresses on the blockchain,
    \item We analyze the security of our consensus mechanism by closely examining the information exchanged on the chain during each stage of the FL process, including the model updates and their associated proofs, and establish through theoretical and empirical analysis that no sensitive information about local models or their training data is disclosed to untrusted parties.
    \item Experimental results demonstrate that our ZKPoT is robust against privacy and Byzantine attacks, as they remain unaffected with $\frac{1}{3}$ malicious clients in the network. ZKPoT achieves such defense performance without the accuracy degradation commonly associated with popular defense mechanisms like DP. Furthermore, the transaction verification and block generation times show minimal increase as the network size scales up.

\end{itemize}

The rest of the paper is organized as follows. We present the related work in Section \ref{sec:related} and the preliminaries in Section \ref{sec:pre}. We detail our method in Section \ref{sec:design}, and present the implementation and evaluation of our method in Section \ref{sec:eval}. Finally, we conclude our paper in Section \ref{sec:conclusion}.

\section{Related Work}\label{sec:related}
In this section, we will introduce the related research, including blockchain consensus protocol and the application of ZKPs in both DL training and blockchain-secured FL 

Building a stable and robust decentralized FL framework crucially depends on selecting an appropriate consensus mechanism, which significantly influences the effectiveness and reliability of the framework. Traditional Proof of Work (PoW) consensus, adopted in networks presented in \cite{Blockchained_on_device, cognitive_com}, is renowned for its security and robust selection process. However, its extensive computational demands render it impractical for most FL scenarios, where massive computational resources are better devoted to model training. Alternative mechanisms like Proof of Stake (PoS) are employed in several frameworks \cite{dpos_iov, pos_iot} to achieve an efficient and energy-conserving selection process. Additionally, Li \textit{et al.} \cite{commitee_consensus} proposed a committee consensus mechanism where a select group of trustworthy clients forms a committee responsible for validating updates from local clients and creating new blocks. In \cite{podl}, the authors introduced Proof of Deep Learning (PoDL) to redirect the computational resources typically used in PoW to DL model training, thereby avoiding the waste of energy on unproductive computations. Similarly, Qu \textit{et al.} \cite{pofl} developed Proof of Federated Learning (PoFL), which tailors the consensus mechanism to better suit FL environments. This approach allows miners to competitively train local models with private data to vie for the leader position, selecting the miner who achieves the highest accuracy in model training as the leader for the current round.

Although these mechanisms offer an energy-recycling advantage, they pose a risk to client privacy during the model validation phase. In this phase, all clients run each others' models to verify claimed accuracy, which may expose their models to unverified nodes. Extensive works \cite{DLG, IDLG, DMIA, MIA} have shown that model parameters can reveal information about the training data, making them susceptible to theft by malicious attackers. To address this privacy limitation, Qu \textit{et al.} \cite{PoQ} and Wang \textit{et al.} \cite{pf-pofl}, who implemented similar model performance-based consensus mechanisms, incorporated differential privacy (DP) to protect model parameters. While this method enhances privacy to some extent, it also introduces significant additional computational demands and generally results in diminished performance. Despite these efforts, the existing solutions still fail to achieve an optimal balance between efficiency and security, leaving a critical gap in the current landscape of blockchain-secured FL.

Zero-knowledge proofs (ZKPs) provide a robust method for verifying model integrity and ensuring the correctness of model computations during inference, without disclosing any model specifics. Ghodsi \textit{et al.} \cite{saftynet} implemented a specialized interactive proof (IP) protocol for verifiable execution of deep neural networks that utilize quadratic activation functions. In \cite{vcnn}, Lee \textit{et al.} developed vCNN, a zero-knowledge succinct non-interactive argument of knowledge (zk-SNARK) scheme specifically designed for convolutional neural networks (CNNs). Furthermore, Feng \textit{et al.} \cite{zen} proposed the ZEN framework to efficiently generate verifiable zero-knowledge inference schemes for CNNs of varying depths and implemented chain coding optimization to reduce the proof overhead. Subsequently, Liu \textit{et al.} \cite{zkcnn} developed zkCNN, underpinned by a novel sumcheck protocol that facilitates the verification of fast Fourier transforms and convolutions, achieving linear prove time and expediting the proof generation process.

Recently, blockchain-secured FL frameworks have increasingly integrated ZKPs to enhance both privacy and robustness. Frameworks such as zkFL \cite{zkfl}, zkFDL \cite{zkfdl}, and ZKP-FL \cite{zkp-fl} utilize ZKPs to secure global model aggregation, with the central server generating proofs that allow every local client to verify the aggregation's correctness. However, these methods have not fully addressed potential privacy leaks that may occur when local model parameters are sent to an unverified central server. Notably, generating a correct proof of aggregation does not preclude the possibility of extracting information from the model parameters. Zhang et al. \cite{zpol} proposed a Zero-Knowledge Proof of Learning (ZPoL) consensus algorithm within their framework, utilizing zkCNN \cite{zkcnn} to verify the performance of local models. Although zkCNN aids in preserving privacy, as an interactive protocol, it requires multiple rounds of communication between the prover and versifier, rendering it less suitable for efficient verification, particularly in decentralized environments. To address the aforementioned limitations, our proposed ZKPoT protects the privacy of local data and models against all unverified parties, while our utilization of a non-interactive ZKP protocol ensures scalable and efficient verification.

\section{Preliminaries}\label{sec:pre}

\subsection{Federated Learning}
Federated learning (FL) is a decentralized machine learning approach, where multiple clients collaboratively train a shared global model without exposing their local data. Each client $i$ trains its own local model $w^{t}_{local, i}$ at the global round $t$ using local data $\mathcal{D}_i$, and periodically shares its model updates with a central server, which aggregates these updates to form a new global model $W^{t+1}_{global}$.

The goal of federated learning is to minimize a global loss function $\mathcal{L}(W)$, which is typically formulated as the weighted average of the local loss functions $\mathcal{L}_i(w)$ on each client's local dataset $\mathcal{D}_i$ \cite{fedavg}:

\[
\min_W \mathcal{L}(W) = \sum_{i=1}^{N} \frac{|\mathcal{D}_i|}{\sum_{j=1}^{N} |\mathcal{D}_j|} \mathcal{L}_i(w^{t}_{local,i}),
\]

where $\mathcal{L}_i(w)$ is the loss function for client $i$, and $N$ is the total number of clients. Each client optimizes its local objective by performing several steps of gradient descent on its local data, updating its local model $w^{t}_{local,i}$. After local training, the clients share their model updates, and the central server aggregates these updates to update the global model:

\[
W^{t+1}_{global} = \sum_{i=1}^{N} \frac{|\mathcal{D}_i|}{\sum_{j=1}^{N} |\mathcal{D}_j|} w^{t}_{local,i}.
\]
The new global model $W^{t+1}_{global}$ is then sent back to the clients for the next round of local training.

\subsection{zk-SNARK}
Zero-Knowledge Succinct Non-Interactive Argument of Knowledge (zk-SNARK) is a form of cryptographic protocol that enables one party to prove possession of specific information (e.g., the solution to a computational problem) without revealing the information itself. A zk-SNARK is defined by a triple of polynomial time algorithms (Setup, Prove, Verify):
\begin{itemize}
    \item $(pk, vk) \leftarrow \text{Setup}(1^\lambda, \mathcal{C}):$ Given a security parameter $\lambda$ and a circuit description $\mathcal{C}$, the $Setup$ function randomly generates a proving key $pk$ and a verification key $vk$.
    \item $(\pi) \leftarrow \text{Prove}(pk, x, w):$ Given a proving key $pk$, public inputs $x$, and a witness $w$, the $Prove$ function computes a proof $\pi$ that attests to the correctness of the computation given $w$, without revealing any information about the witness.
    \item $\{0, 1\} \leftarrow \text{Verify}(vk, x, \pi):$ Given a verification key $vk$, public inputs $x$, and a proof $\pi$, the $\text{Verify}$ function outputs 1 if $\pi$ is a valid proof for the inputs under $vk$, and 0 otherwise.
\end{itemize}
This structured approach allows zk-SNARKs to satisfy several critical cryptographic conditions, which we detail as follows:
\begin{itemize}
    \item \textit{Completeness:} Given any security parameter $\lambda$, any circuit $C$, and true statement $x$, an honest prover can convince the verifier. Formally, this is expressed as:
    \begin{equation}
\Pr\left[\text{Verify}(vk, x, \pi) = 1 \, \middle| \,
    \begin{aligned}
    &(pk, vk) \leftarrow \text{Setup}(1^\lambda, \mathcal{C}),\\
    &\pi \leftarrow \text{Prove}(pk, x, w)
    \end{aligned}
    \right] = 1
\end{equation}
    \item \textit{Succinctness:} The length of a proof satisfies
    \begin{equation}
    |\pi| \leq \text{poly}(\lambda)\text{polylog}(|x| + |w|)
    \end{equation}
    \item \textit{Proof of Knowledge:} A prover capable of generating valid proof must indeed possess the corresponding witness to the statement being proven. Formally, for an adversary without the corresponding witness, the probability of constructing a valid proof is denoted as:
    \begin{equation}
    \Pr\left[
    \begin{aligned}
    &(pk, vk) \leftarrow \text{Setup}(1^\lambda, \mathcal{C}); \\
    &\pi \leftarrow \text{Prove}(pk, x, w) : \\
    &w \notin \mathcal{W} \text{ and } \text{Verify}(vk, x, \pi) = 1
    \end{aligned}
    \right] \approx 0.
    \end{equation}
    \item \textit{Zero Knowledge:} A proof reveals nothing but the truth of the statement. Formally, for every verifier, there exists a simulator $Sim$ that can generate a proof that is indistinguishable from an actual proof, without having access to the witness. It is expressed as:
    \begin{align}
    \Pr\left[
    \begin{aligned}
    &(pk, vk) \leftarrow \text{Setup}(1^\lambda, \mathcal{C}),\\ 
    &\pi \leftarrow \text{Prove}(pk, x, w)
    \end{aligned}
    \, \middle| \, \mathcal{V}(vk, x, \pi) = 1 \right] & \nonumber \\
    = \Pr\left[
    \begin{aligned}
    &(pk, vk) \leftarrow \text{Setup}(1^\lambda, \mathcal{C}),\\ 
    &\pi \leftarrow \text{Sim}(1^\lambda, vk, x)
    \end{aligned}
    \, \middle| \, \mathcal{V}(vk, x, \pi) = 1 \right] &.
\end{align}
\end{itemize}

\subsection{IPFS}
The InterPlanetary File System (IPFS) is a decentralized protocol designed to enhance the efficiency and security of data storage by using content-based addressing through cryptographic hashes \cite{ipfs}. This approach allows IPFS to ensure data integrity and eliminate redundancies across the network, as each piece of data is uniquely identified by its hash. Additionally, IPFS employs a Distributed Hash Table (DHT) for dynamic content discovery and routing, further enhancing data retrieval efficiency. We use IPFS to store the global model ZK proofs, reducing the storage cost of the blockchain and ensuring the integrity of the global model.

\section{Zero-knowledge Proof of Training (ZKPoT)}\label{sec:design}
\begin{figure*}[htbp]
    \centering
    \includegraphics[width=0.8\textwidth]{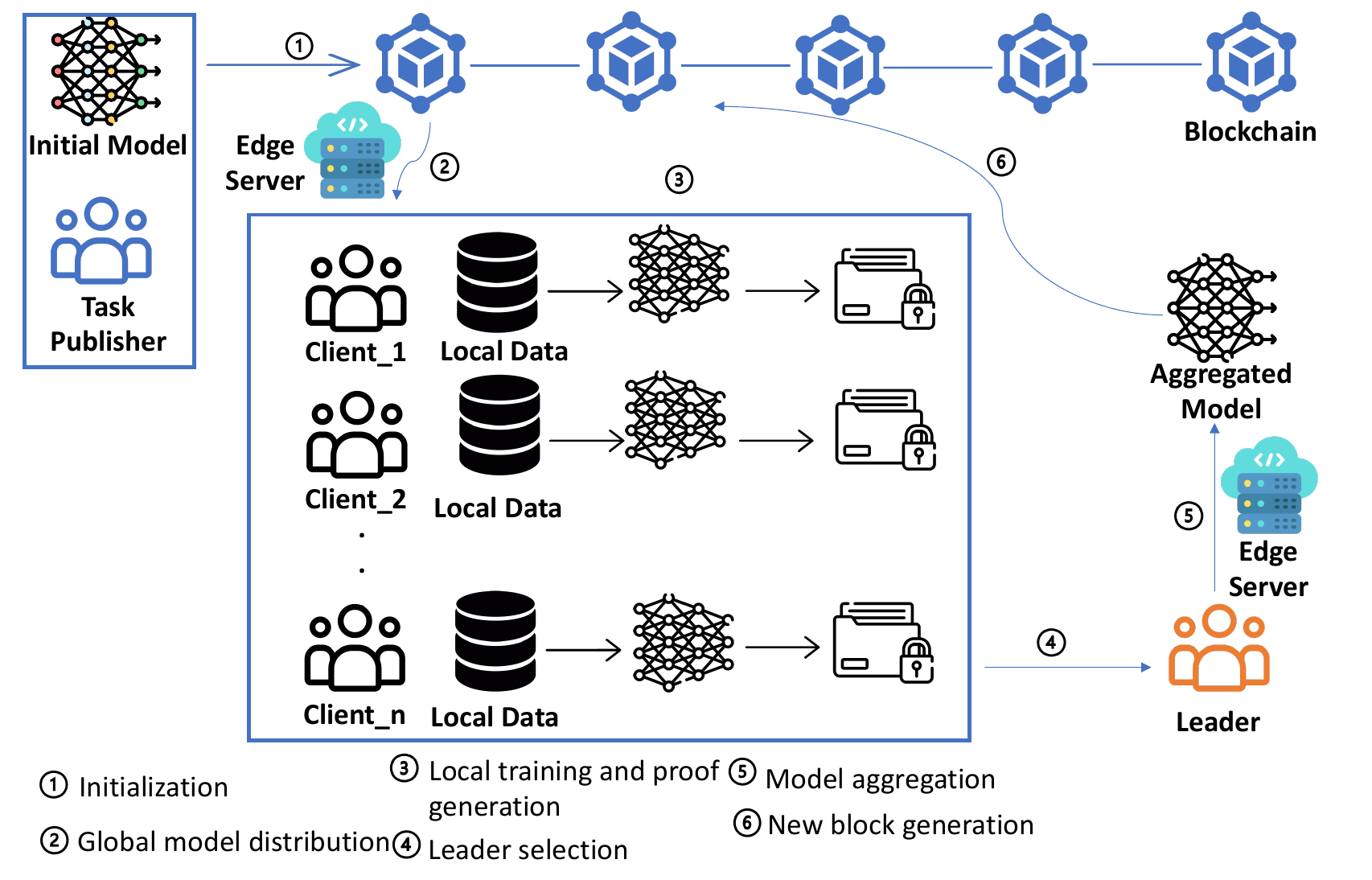}
    \caption{The architecture of ZKPoT-based Blockchain-secured Federated Learning system}
    \label{fig:workflow}
\end{figure*}
In this section, we propose a blockchain-secured FL framework with ZKPoT consensus mechanism. With this consensus, we hope to achieve the following goals: 1) establish a decentralized training environment that circumvents the security vulnerabilities associated with a central server; 2) safeguard the privacy of local data and models without compromising the performance of the global model; 3) achieve an efficient and robust leader selection process while reducing the computational cost for clients; 4) ensure scalability to accommodate a large number of blockchain nodes.

\textbf{Design Overview:} As shown in Fig.\ref{fig:workflow}, the task publisher first initializes the blockchain network, publishes the FL task, and uploads the IPFS addresses for the initial global model (Step \textcircled{1}). Interested clients register with the blockchain and download the global model from the most recent block to start local training with their private data (Step \textcircled{2}). After completing local training, each client computes the local model's accuracy $acc_i$ using a test dataset given by the task publisher, and generates a corresponding proof $\pi_{acc_i}$ (Step \textcircled{3}). The client then sends a transaction to the transaction pool that includes the IPFS address of both the local model and the ZK proof, along with other necessary verification information. The transactions are sorted by accuracy, and nodes begin the verification starting from the highest. This process continues until proof is verified as valid, determining the leader for that training round (Step \textcircled{4}). The selected leader then verifies the remaining proofs to ensure they are valid and that the reported accuracy meets the required threshold. Upon successful verification, the leader aggregates the approved local models into a new global model (Step \textcircled{5}). The leader then creates a new block, which includes the IPFS address of this aggregated global model and the Merkle root of the transactions related to the local models that contributed to the aggregation. This block is subsequently added to the blockchain and broadcast to ensure all network participants can synchronize with the updated state (Step \textcircled{6}).

\begin{table}[h]
\caption{List of Notations}
\centering
\begin{tabularx}{\linewidth}{|l|X|}
\hline
\textbf{Notation} & \textbf{Description} \\
\hline
$W^{t}_{global}$ & Global model constructed at the end of global round $t$ \\
\hline
$w^{t}_{locla}$ & Local model trained at global round $t$  \\
\hline
$G_t$ & IPFS address of global model $W^{t}_{global}$ \\
\hline
$\mathcal{Q}$ & Quantized model \\
\hline
$cm_i$ & Commitment from client $i$ \\
\hline
$acc_i$ & Accuracy calculated by client $i$ on the test dataset \\
\hline
$\pi_{acc_i}$ & Zero-knowledge proof on $acc_i$ \\
\hline
$pk$ & Proving key   \\
\hline
$vk$  &  Verification key    \\
\hline
$D$ & Test dataset \\
\hline
$T$ & Truth labels \\
\hline
\end{tabularx}
\label{tab:notations}
\end{table}

\subsection{Initialization}
We consider the task publisher to be ``semi-honest", meaning it reliably performs essential functions such as participating in the trusted setup phase of our zk-SNARK protocol, which requires a trusted third party. However, we do not dismiss the possibility that the task publisher may be ``curious" about sensitive information, specifically clients' model parameters \cite{jebreel2022enhanced}. Therefore, the task publisher does not have direct access to local model parameters unless it participates in the mining process and is elected as the leader node.

In the initialization phase, the task publisher creates the genesis block of the blockchain network. This block includes the FL task, the IPFS address $G_0$ of the initial global model parameters $W_{global}^0$, and the hyperparameters necessary for training. To prevent clients from training directly on them and to ensure unbiased model development, the test datasets are not released at this stage. Interested clients register on the blockchain network and prepare for the training process.

\subsection{Transaction and Block Design}
To ensure compatibility with our system design and to meet its unique requirements, we design a new block and transaction format. A blockchain comprises an ordered series of blocks, with each one securely linked to its successor through the incorporation of the previous block's hash in its header—this mechanism reinforces the integrity of the chain \cite{zhu2023blockchain}. Within ZKPoT, the block header encapsulates the block's sequence number, the timestamp of creation, and the hash of the antecedent block. The body of each block is bifurcated into transactions and the IPFS address where the global model is stored. To optimize network functionality, we introduce a novel transaction format that encapsulates several critical elements: the sender's identifier, the designated task name, the model's accuracy, the IPFS location of both the model and its associated proof, as well as the proving key and the verification key. After proof generation, participants dispatch transactions containing their models to the transaction pool, ensuring model immutability during the subsequent verification phase. 

\subsection{Distributed Training}
After initialization, registered clients retrieve the initial global model for training. In our system, clients identify the most recent block by its height, the highest number in the blockchain sequence, to obtain the IPFS address for the latest global model $W_{global}^{t-1}$ from the last block on the blockchain. With $W_{global}^{t-1}$ and their private data, clients compute new local models $w_{local}^{t}$.

\subsection{Quantization}
Because zk-SNARK systems operate exclusively within finite fields where data is represented as integers, clients need to quantize their trained models -- convert floating-point numbers into integers before generating zk-SNARK proofs. The quantization scheme is an \textit{affine mapping} of integers $q$ to real numbers $r$, i.e. of the form $r = S(q-Z)$ for some scale parameters S and zero points Z. To facilitate inference with quantized models, a corresponding computational framework for integer parameters has been implemented. Considering floating-point matrices $W$ and $X$, a neural network computes the output matrix $Y$ as: 
\begin{equation}
Y=WX, W\in \mathbb{R}^{m\times n}, X\in \mathbb{R}^{n}, Y\in \mathbb{R}^{m}.
\label{one}
\end{equation}
Initially, we determine scale parameters ($s_Y, s_W, s_X \in \mathbb{R}$) and zero points ($z_Y, z_W, z_X \in \mathbb{Z}$), thereby attaining a quantized representation for each matrix:
\begin{align*}
Y = s_Y(Q_Y - z_YJ_{m,1}), && W = s_W(Q_W - z_WJ_{m,n}),
\end{align*}
\[
X = s_X(Q_X - z_XJ_{n,1}),
\]
where $J_{k, l}$ represents a $k\times l$ matrix of ones. Upon replacing $Y$, $X$, and $W$ with their quantized counterparts, we arrive at the equation:
\begin{equation}
s_Y(Q_Y - z_YJ_{m,1}) = s_W(Q_W - z_WJ_{m,n})(Q_X - z_XJ_{n,1}).
\end{equation}
Then we replace the floating-point scale parameters utilizing a multiplier $M$ to enable full quantization computation:
\[
M = 2^k \frac{s_W \cdot s_X}{s_Y}, 
\]
\begin{equation}
Q_Y - z_Y J_{m,1} = M(Q_W - z_W J_{m,n})(Q_X - z_X J_{n,1}) / 2^k.
\label{Three}
\end{equation}

To accommodate computations within a finite field, our approach incorporates two R1CS-optimized techniques—sign-bit grouping and remainder-based verification, as expounded in \cite{zen}. These methods are specifically adopted to curtail the frequency of computationally intensive operations inherent in the generation and verification of ZKPs, particularly bit decomposition and division. With sign-bit grouping, we reconstruct \eqref{Three} to guarantee the positivity of all intermediate values, thereby obviating the necessity for sign check and eliminating the requirement for bit-decomposition:
\begin{align*}
    G_1 &= QWQX,  G_2 = z_xQW,  G_3 = zwQX, 
    M' = \left\lceil\frac{z_y2^k}{M} \right\rceil,
\end{align*}
\begin{equation}
QY = M(G_1 + nz_wz_xJ_{m,1} + M'J_{m,1} - G_2 - G_3)/2^k.
\label{four}
\end{equation}
We proceed by implementing remainder-based verification, streamlining the verification process for division operations, which are prevalent in average pooling kernels and quantization stages. This method reformulates the matrix multiplication as follows:
\begin{equation}
QY2^{k} + R = M(G_{1} + nz_{w}z_{x}J_{m,1} + M'J_{m,1} - G_{2} - G_{3}),
\label{five}
\end{equation}
where R is an extra matrix to store the division remainder.

\subsection{Zero-knowledge Model Verification}
Upon completing quantization, clients proceed to the proving phase and generate proofs to demonstrate their models' performance. To ensure the security and privacy of local models, our ZKPoT leverages a typical type of zk-SNARK system, namely Groth16 \cite{groth16}, to allow clients to generate proof that verifies that their models achieve the claimed accuracy on the public dataset without disclosing any model details.

The security and reliability of a zk-SNARK system are fundamentally dependent on the setup phase, during which a common reference string (CRS) that contains the proving key $pk$ and the verifying key $vk$ are generated. This CRS, generated through a randomized process, is tailored to a specific program— in our context, this program corresponds to a particular type of DL model. In our design, the task publisher is designated as a trusted third party and is responsible for conducting the setup phase using the global model. This setup occurs simultaneously with the local training processes of other clients, ensuring efficient integration and continuity of operations. 

\begin{figure*}[t]
    \centering
    \includegraphics[width=0.8\textwidth]{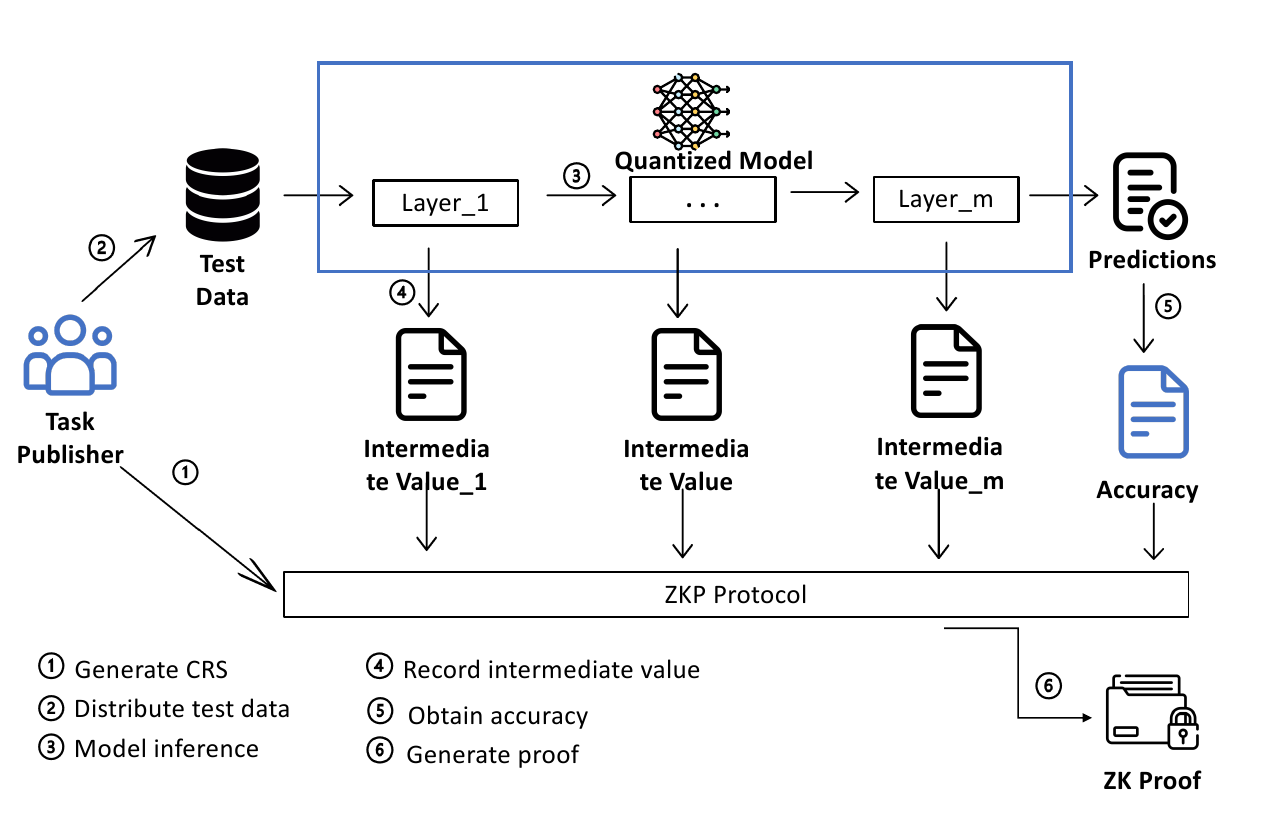}
    \caption{Zero-knowledge proof generation process}
    \label{fig:zk process}
\end{figure*}

After a client completes its local training and quantization, it needs to commit the quantized model $Q$ into a commitment $cm$. A commitment guarantees two properties: hiding and binding. The hiding property ensures that no information about the model $Q$ is leaked from the commitment, while the binding property ensures that the committed model cannot be altered without detection. Specifically, we leverage the Pedersen commitment scheme to compute clients' commitments on their models. Given a quantized local update $w_i$, a client computes the commitment as follows: $$cm = g^{w_i} \cdot h^{r_i}$$ where $g$, $h$ are public group generators, and $r_i$ is a randomness generated by the client. The client initially transmits the commitment $cm$ to the task publisher. Upon validating the commitment, the task publisher dispatches the testing dataset $D$ along with the associated truth labels $T$ back to the client. Subsequently, the client executes the inference process on the dataset $D$ utilizing the computational framework for integer parameters outlined in the quantization section. As illustrated in Fig. \ref{fig:zk process}, it is necessary to document all intermediate values generated during computations in each layer for proof generation. For each layer, these intermediate values comprise the weights and outputs, which are recorded as integer parameters $q$, scale parameters $s$, and zero points $z$. After completing the inference computations, the client obtains the predictions $Y$ and compares them with the truth labels $T$ to determine the accuracy, denoted by $acc$, of the committed model. After evaluating the model's accuracy, the client begins the proof generation process by translating the recorded computation details into a format suitable for cryptographic verification. This is achieved by converting the computations into a Rank-1 Constraint System (R1CS), which is a standard approach for expressing computations in zk-SNARKs. The R1CS format encapsulates the operations performed during inference into a series of linear constraints. Using $pk$ provided by the task publisher, the client then constructs a cryptographic proof that demonstrates the computations satisfy these constraints. With this proof, any client within the blockchain network can use the verification key $vk$, which is publicly accessible on the network, to confirm the accuracy of the committed model.
Based on these steps, we define the zero-knowledge proof scheme of ZKPoT as the following algorithms:
\begin{itemize}
    \item $(pk, vk) \leftarrow \textit{ZKPoT.Setup}(1^\lambda, \mathcal{Q})$: Task publisher randomly generates a proving key $pk$ and a verification key $vk$, using a security parameter $\lambda$ and the global model $\mathcal{Q}$.
    \item $cm \leftarrow \textit{ZKPoT.Commit}(\mathcal{Q}, r)$: Given a random opening r, the prover commits to its local model $\mathcal{Q}$, and outputs a commitment $cm$.
    \item $(acc, \pi_{acc}) \leftarrow \textit{ZKPoT.Prove}(pk, D, T, \mathcal{Q})$: Prover runs inference computation for each data sample in $D$, compares the predictions with $T$ to obtain $acc$, and generates the corresponding proof $\pi_{acc}$.
    \item $\{0, 1\} \leftarrow \textit{ZKPoT.Verify}(vk, cm, D, T, acc, \pi_{acc})$: Given the verification key $vk$ and proof $\pi_{acc}$, verifies if the following statements are true: $cm$ is a commitment to $\mathcal{Q}$; the number of correct predictions on dataset $D$ using model $\mathcal{Q}$ is $acc$. 
\end{itemize}
\subsection{Security Analysis}
In this section, we analyze the security of ZKPoT, with a specific focus on its ability to prevent information leakage concerning model parameters. In the setup of our Groth16 zk-SNARK, the task publisher runs \textit{ZKPoT.Setup()} function that utilizes the security parameter \( \lambda \) to determine the cryptographic strength of the system, which dictates the selection of elliptic curve groups and the bit lengths of field elements to provide the desired level of security against cryptographic attacks. The CRS is generated independently of the specific parameters of the computational problem \( \mathcal{Q} \), yet it is tailored to the type and size of \(\mathcal{Q} \)'s circuit. Generators \( g_1 \) and \( g_2 \) for groups \( G_1 \) and \( G_2 \) are selected without any information about \( \mathcal{Q} \)'s specifics, ensuring that no sensitive data is embedded within these elements. Additionally, scalars such as \( \alpha, \beta, \gamma, \delta \) are randomly generated from the underlying field, influenced solely by \( \lambda \) and not by any specific aspects of \( \mathcal{Q} \). This method of CRS construction secures the proof system while maintaining independence from \( \mathcal{Q} \)'s internal parameters, thereby ensuring that the CRS supports the integrity and zero-knowledge properties of the proofs related to \( \mathcal{Q} \) without revealing any proprietary or sensitive information through the setup process. 

Before generating proofs, clients commit to their local models and later share that commitment for verification. Specifically, a client commits the parameters of its local model \( w_i \) as \( cm = g^{w_i} h^r_{i} \), where the randomness introduced by \( r \), chosen uniformly from a large space, obscures \( w_i \). This ensures that the commitment \( cm \) appears uniformly random within the group generated by \( g \) and \( h \). Since \( h \) is selected such that its discrete logarithm relative to \( g \) (i.e., \( \log_g h \)) is unknown, it is computationally infeasible for an adversary to extract \( v \) or \( r \) from the commitment. Thus, the commitment effectively hides the model parameters, ensuring that they do not leak to unauthorized parties.

\begin{algorithm}[t]
\caption{ZKPoT Consensus}
\label{alg:ZKPoT_consensus}
\begin{algorithmic}[1] % The number [1] here enables line numbering.
\\State Initialize transaction counter $n \gets 0$
\State Determine total transactions expected $N$
\State Define the structure of each transaction as \textit{tx} = (\textit{Model}, \textit{sender\_id}, \textit{acc}, \textit{$\pi_{acc}$}, \textit{cm});
\State Initialize transaction counter $n \gets 0$;
\State Determine total transactions expected $N$
\While{Current Time $\leq$ Deadline and $n < N$} % Loop condition checking if the current time is before the deadline and all transactions are not yet received
    \State Wait for incoming transaction % Action to take if within the loop
    \If{transaction received}
        \State Add received transaction to the transaction pool % Action to add transaction to the pool
        \State $n \gets n + 1$ % Increment the transaction counter
    \EndIf
\EndWhile
\State Sort(transactions, acc, descending)
\For{each \textit{tx} in \textit{sorted transaction pool}}
    \State Initialize vote count $votes \gets 0$
    \For{each client in network}
        \If{\Call{Verify}{\textit{tx}}}
            \State $votes \gets votes + 1$
        \EndIf
    \EndFor
    \If{$votes \geq \frac{2N}{3}$}
        \State \textbf{break} and declare \textit{sender\_id} from \textit{tx} as the leader;
    \EndIf
\EndFor
\State End the verification and leader election process.
\end{algorithmic}
\end{algorithm}

During the proving and verifying processes, the validity and security of the model parameters are ensured by our \textit{ZKPoT.Prove()} and \textit{ZKPoT.Verify()} algorithms. Specifically, \textit{ZKPoT.Prove()}, adhering to the proof of knowledge property of zk-SNARKs, requires that the prover possesses a model \( \mathcal{Q} \) that achieves accuracy \( acc \) on dataset \( D \) in order to generate the proof \( \pi_{acc} \). On the other hand, the zero-knowledge property ensures that the proof \( \pi_{acc} \) reveals nothing beyond the veracity of the statement. Therefore, all steps involved in our model verification process ensure that no information regarding the model parameters is disclosed.

\subsection{Consensus Achievement}
After generating a proof, a client creates a transaction that includes the following components: the IPFS location of its model $\mathcal{Q}$, the corresponding proof $\pi_{acc}$, the accuracy $acc$, and the commitment $cm$. The transaction is then submitted to the transaction pool, which continues to accept transactions until either a predefined deadline is reached or all participants have submitted their transactions. Upon ceasing to receive new transactions, the task publisher sorts the transactions within the pool based on their accuracy. Subsequently, all clients in the network participate in verifying each transaction using the \textit{ZKPoT.Verify()} algorithm we defined earlier. Transactions are sorted by performance, so the first one that passes verification—achieving ``Yes" votes from two-thirds of the total number of clients—is considered the highest-performing and is therefore selected as the leader. The leader then verifies the remaining transactions and initiates a new transaction containing the IPFS locations of all verified models. Next, the leader conducts the model aggregation process, utilizing FedAvg \cite{fedavg}, to derive the new global model $W^{i}_{global}$. Finally, the leader compiles a new block with the verified transactions and the IPFS location of the new global model, appending this block to the blockchain.

\section{Performance Evaluation}\label{sec:eval}

In this section, we evaluate the performance of the proposed framework with the ZKPoT consensus mechanism. 

\subsection{Experimental Setup}

\subsubsection{System Settings}

We implemented ZKPoT with Python 3.10 and Rust 1.77. The networking and distributed aspects of our design are built with Python. We utilized Pytorch 2.2 to train deep-learning models. Our ZKP protocol was implemented using Rust, with Groth16 \cite{groth16} as the underlying zk-SNARK scheme. We adopted the elliptic curve BLS12-381 implementation from the ark-bls12-381 library for cryptographic operations. For the interface between Python and Rust, we utilized the PyO3 library. We deployed ZKPoT on a machine equipped with a single NVIDIA GeForce RTX 4070 Laptop GPU and a 13th Generation Intel(R) Core(TM) i7-13800H CPU, operating at 2.50 GHz. By default, our experiments simulate a blockchain network consisting of 100 nodes, representing the participants involved in the FL task.

\subsubsection{Datasets and Machine Learning Models}
All experiments were executed on two datasets: CIFAR10 and MNIST. The CIFAR-10 dataset comprises 60,000 color images, each measuring $32\times32$ pixels, organized across 10 classes with 6,000 images per class, and divided into 50,000 training images and 10,000 test images. The MNIST dataset features 70,000 images of handwritten digits 0-9, normalized and centered within a $28\times28$ pixel frame, split into 60,000 training and 10,000 test images. In the FL setup, 100 clients are loaded, with 20 selected for participation in each training round. For each FL task, we perform 200 rounds of global training. In each round, the selected clients perform 5 epochs of local training with their local dataset. In our experiments, we use an adapted LeNet model for the CIFAR10 dataset. Our configuration starts with two convolutional layers, each followed by ReLU activation and average pooling, suitable for handling both $32\times32$ and $28\times28$ input images. The first convolutional layer contains 6 filters of size 5$\times$5, while the second layer expands to 16 filters of the same size. A third convolutional layer with 120 filters of size $4\times4$ captures finer details. The network concludes with two linear layers, scaling down from either 480 or 120 to 84 neurons, and finally to 10 output neurons for class predictions, with ReLU activations throughout. Additionally, for the MNIST data, we utilize a ShallowNet with two fully connected layers with shapes (784, 128) and (128, 10), respectively, and a ReLU activation function between them.

\subsubsection{Evaluation Metrics} To assess the effectiveness of our method, we evaluate several key aspects: the stability and accuracy of our system models, their resilience against privacy breaches, and their robustness to Byzantine attacks. We selected these metrics for the following reasons. First, although alternative defense mechanisms such as DP offer protection against privacy attacks, they frequently involve trade-offs that may affect model performance and stability. Consequently, it is crucial to evaluate our approach to ensure it safeguards privacy without sacrificing the other two critical aspects of system performance — stability and accuracy. Secondly, building on this foundation, we assess our approach's resistance to the most prevalent privacy attacks, specifically membership inference attacks and model inversion attacks. Lastly, we demonstrate the security of our design by confirming its resilience to Byzantine faults.

\subsection{Performance Analysis of ZKPoT}
\begin{figure}[t]
\vspace{-0.8cm}
    \centering
    \includegraphics[width=\linewidth]{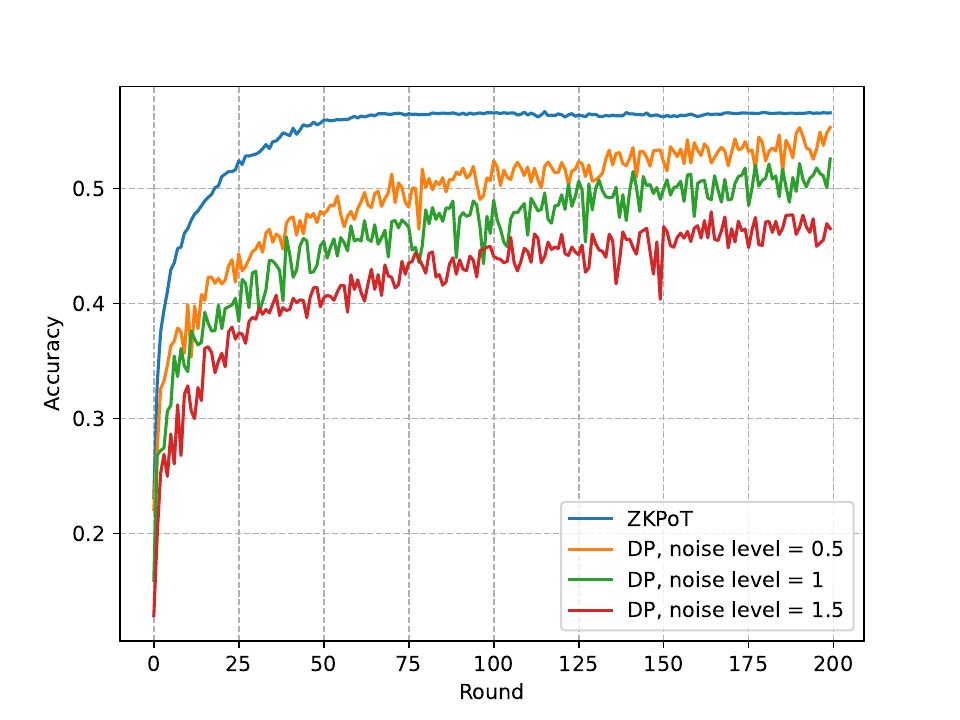}
    \caption{Global model accuracy comparing our method to DP methods with different noise levels on the CIFAR10 dataset}
    \label{fig:noise-1}
\end{figure}

In this section, we analyze the global model performance within our design. We compare our approach with DP, which is the major defense mechanism used by most model performance-based consensus mechanisms, such as \cite{PoQ, Blockchained_on_device}. All experiments are conducted under consistent FL settings, with the distinction that our approach protects the local models using ZKPoT. In contrast, other approaches utilize DP and add Gaussian noise to local training steps. We add Gaussian noise  using the Opacus library \cite{opacus} and fine-tune the hyperparameters to control the noise level.
As shown in Fig. \ref{fig:noise-1}, ZKPoT consistently achieves the highest accuracy, exceeding 0.5 and demonstrating stability post-convergence. While the model with a noise level of 0.5 approaches the accuracy of our approach, it exhibits less stability, suggesting a compromise in performance consistency. Models with higher noise levels of 1 and 1.5 show progressively lower accuracies and slower convergence speeds, underlining the significant impact of increased noise on learning efficacy in FL environments. On the other hand, with the MNIST dataset, although all models demonstrate higher stability due to simpler models used for training, our approach surpasses the performance of the others.

\begin{figure}[t]
\vspace{-0.5cm}
    \centering
    \includegraphics[width=\linewidth]{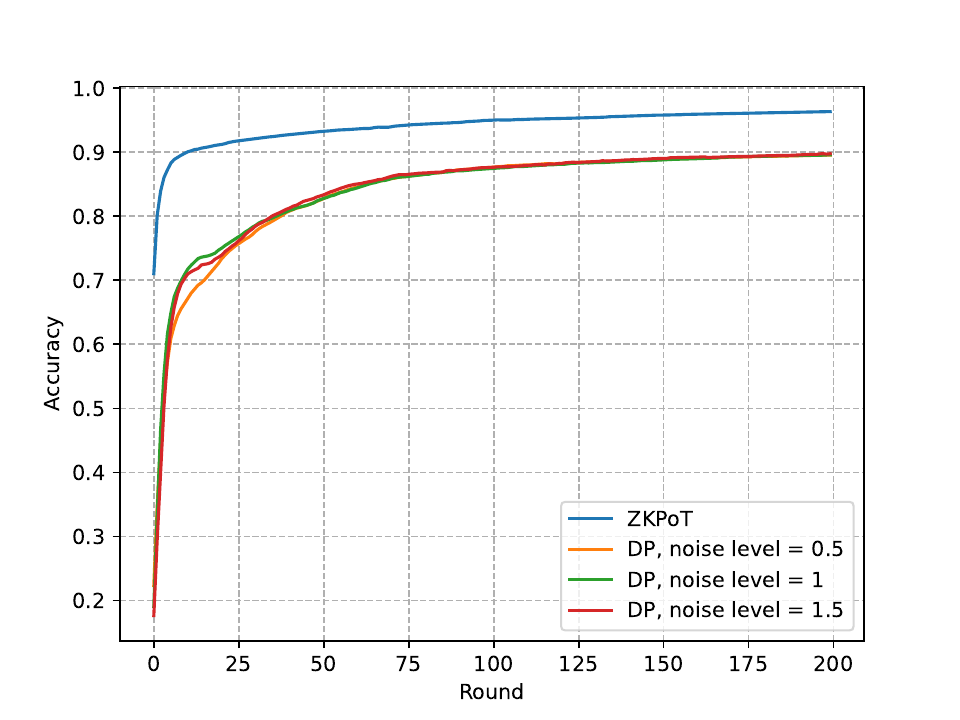}
    \caption{Global model accuracy comparing our method to DP methods with different noise levels on the MNIST dataset}
    \label{fig:noise-2}
\end{figure}

\subsection{Privacy Analysis of ZKPoT}
In this section, we evaluate the robustness of our approach against two prevalent types of privacy attacks: the membership inference attack and the model inversion attack. 
\subsubsection{Membership Inference Attack}
\begin{figure}[t]
\vspace{-0.8cm}
    \centering
    \includegraphics[width=\linewidth]{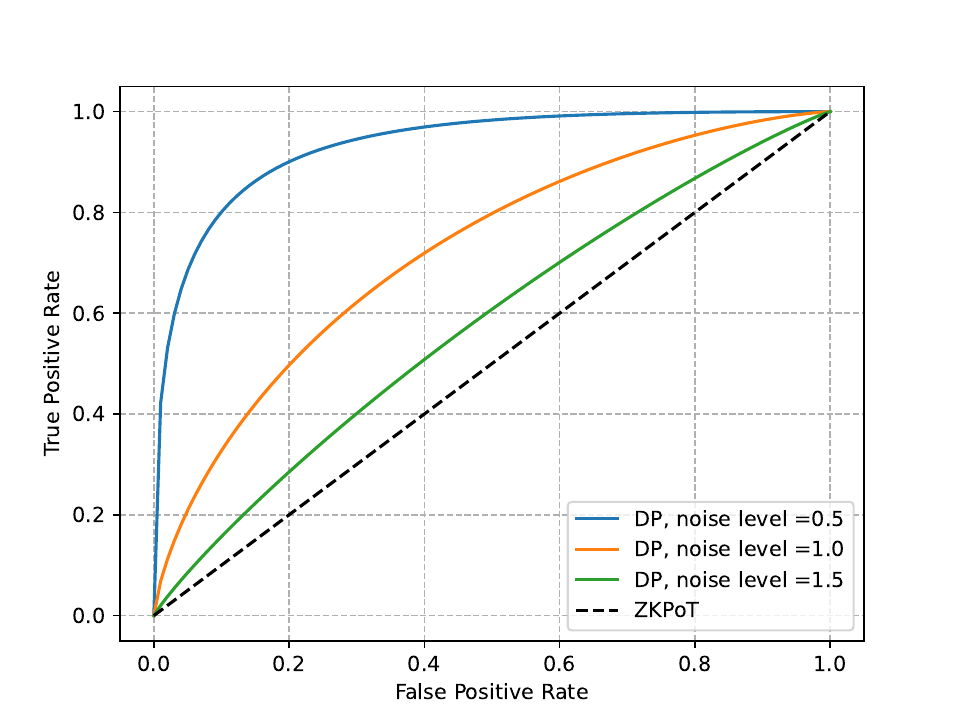}
    \caption{Effectiveness of membership inference attack during the model verification phase on the CIFAR10 dataset. ZKPoT conceals model parameters from unverified clients, reducing the attack to near-random guessing}
    \label{fig:member-1}
\end{figure}

The target of the membership inference attack is to create a binary classifier that determines whether or not a given data sample is within the original training data. 
In the attack method we used, attackers leverage the distribution of the target model gradients to approximate the distribution of the training data \cite{gaussianmip}. We conducted this experiment using the same settings described in the previous section but with the inclusion of at least one malicious client participating in the training each round. The malicious client attempts a membership inference attack on another local model during the model verification phase. The attack’s effectiveness was evaluated by measuring the true positive rate of the binary classifier in identifying data samples from the original training data.  As illustrated in Fig. \ref{fig:member-1}, while adding Gaussian noise provides some level of protection by limiting the true positive rate of the attacker's estimation on the data distribution, our approach significantly enhances security. By completely concealing model parameters from unverified clients, our method ensures that any attack attempting to model the local training data would be reduced to mere random guessing.

\begin{figure}[t]
\vspace{-0.3cm}
    \centering
    \includegraphics[width=\linewidth]{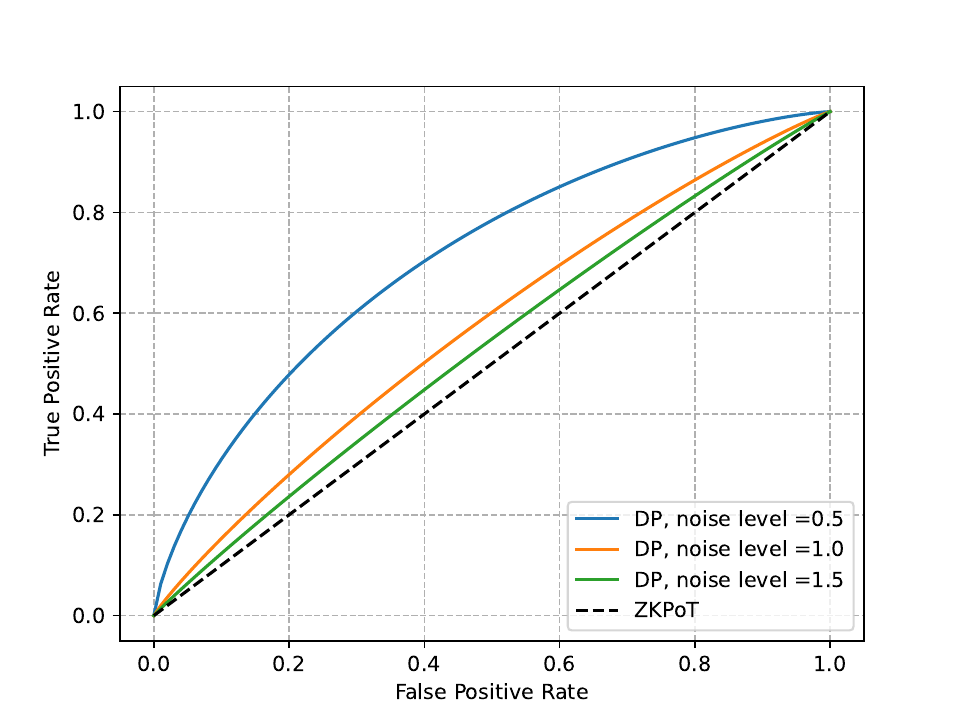}
    \caption{Effectiveness of membership inference attack during the model verification phase on the MNIST dataset}
    \label{fig:member-2}
\end{figure}
\subsubsection{Model Inversion Attack}
\begin{figure}[t]
\vspace{-0.9cm}
    \centering
    \includegraphics[width=\linewidth]{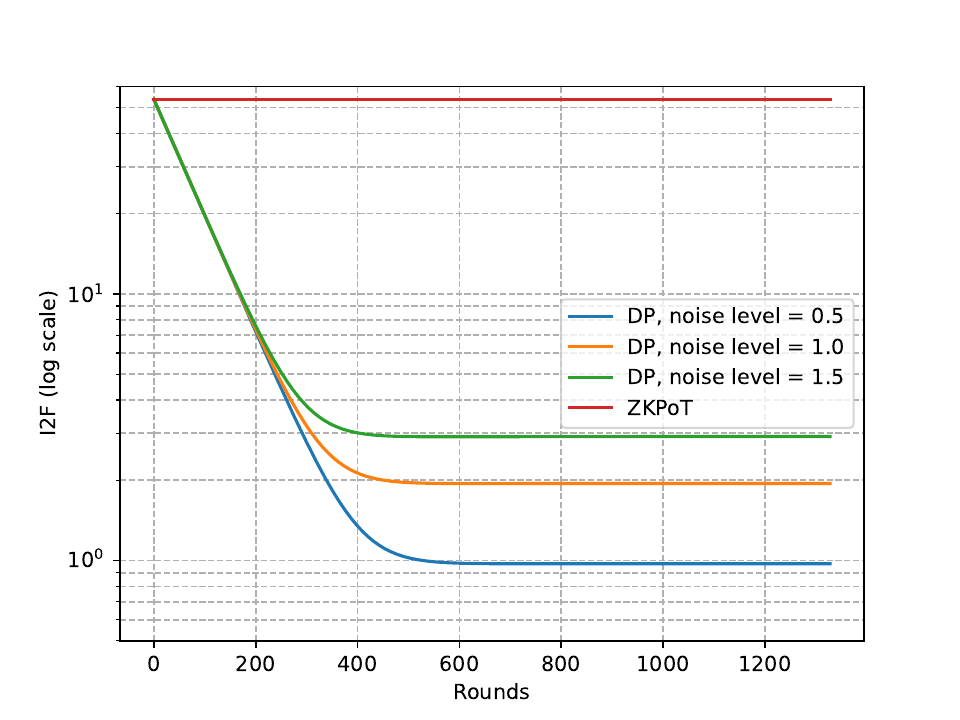}
    \caption{Quality of data reconstruction during model verification phase measured by the inversion influence function on the CIFAR10 dataset}
    \label{fig:i2f}
\end{figure}

\begin{figure}[t]
    \centering
    \includegraphics[width=\linewidth]{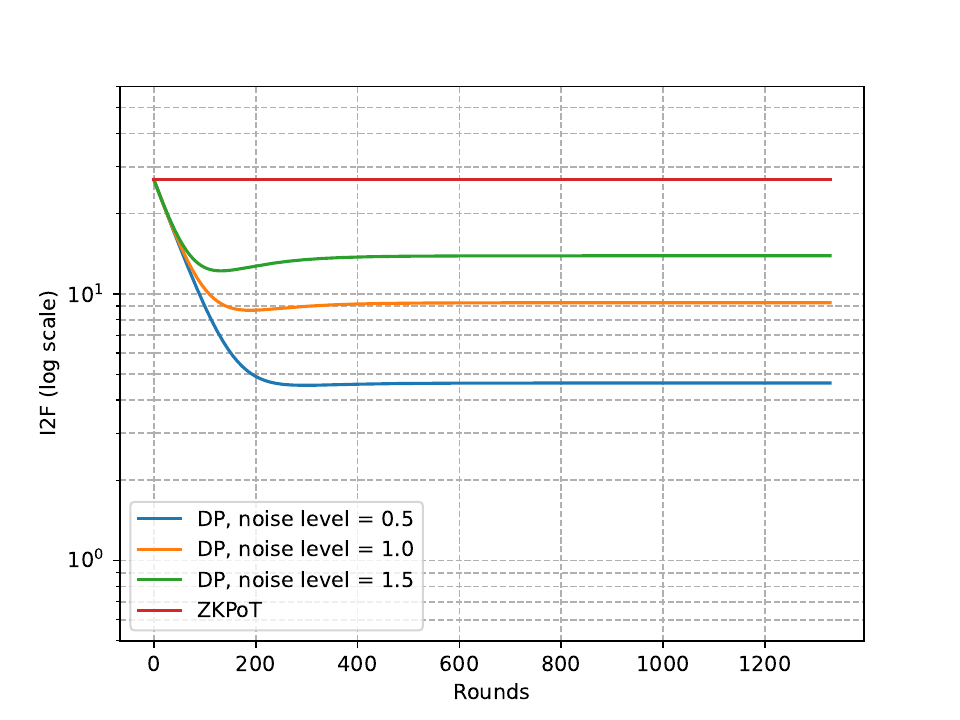}
    \caption{Quality of data reconstruction during model verification phase measured by the inversion influence function on the MNIST dataset}
    \label{fig:i2f mnist}
\end{figure}

In model inversion attacks, attackers attempt to reconstruct the training dataset by exploiting the target model's parameters through techniques like gradient leakage \cite{DLG} and numerical reconstruction \cite{geiping2020inverting}. These attacks can occur in either white-box or black-box scenarios. In white-box attacks, the attackers have direct access to the target model's parameters. Conversely, in black-box scenarios, they do not have direct parameter access and can only interact with the model through inference queries. In our experiment, we allow the malicious client to conduct a white-box numerical reconstruction on another client's model. The attacker initiates the attack with a randomly generated dummy dataset and optimizes it in each iteration of the attack to maximize the cosine similarity between the synthesized gradient and the ground-truth gradient. To evaluate the quality of the reconstructed data, we employ a metric known as the inversion influence function ($I^{2}F$), as detailed in \cite{i2f}. This metric quantifies the semantic distance between the recovered images and the original images. A lower $I^{2}F$ score indicates more accurate reconstruction, while a higher score suggests poorer quality. As depicted in Figs. \ref{fig:i2f} and \ref{fig:i2f mnist}, across both models and datasets, introducing additional noise to the local training effectively impedes data reconstruction, resulting in higher $I^{2}F$ values. However, our approach completely restricts unverified clients from accessing other local models, whether in a white-box or black-box context, thereby preventing the attacker from optimizing the dummy data. 

\subsection{Security Analysis of ZKPoT}
\begin{figure}[t]
\vspace{-0.9cm}
    \centering
    \includegraphics[width=\linewidth]{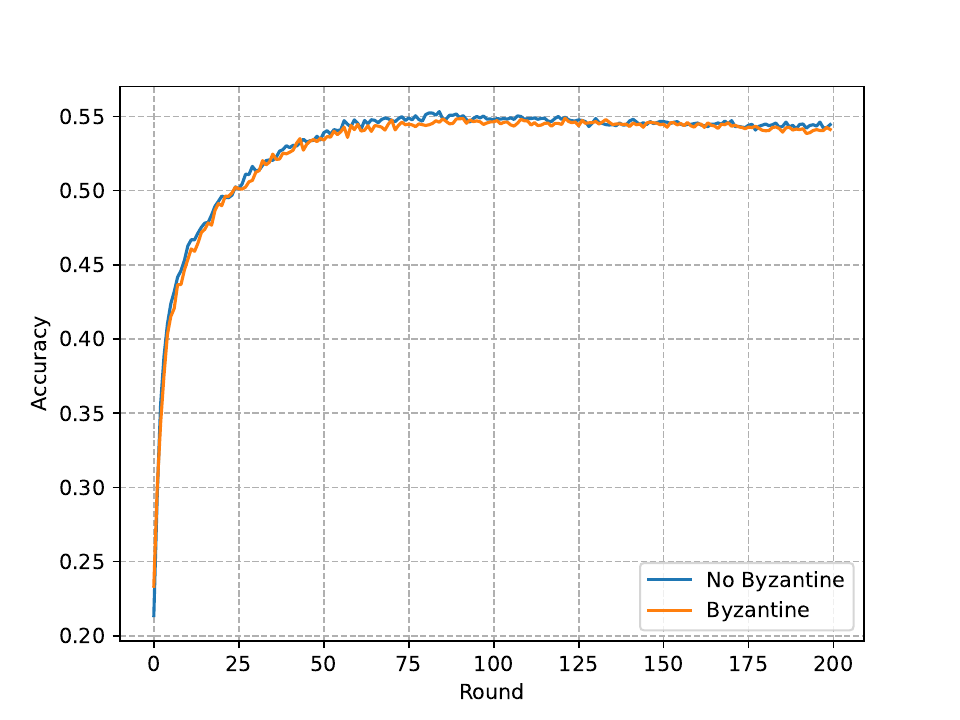}
    \caption{Test accuracy of ZKPoT under honest client settings for CIFAR10}
    \label{fig:byzantine}
\end{figure}

In this section, we evaluate the security of our system, which is protected by the ZKPoT consensus protocol. We assess the resilience of ZKPoT against Byzantine blockchain nodes using the following attack strategy: the malicious client aims to dishonestly become the selected leader. Given that training, a competent model demands substantial computational resources and extensive training data, coupled with the uncertainty of surpassing all other local models, a malicious client may falsely claim an accuracy higher than its actual performance. Once selected as the leader, rather than honestly aggregating local model updates, a malicious client executes a Gaussian attack by assigning random values that follow the standard Gaussian distribution to the global model parameters. We assign one-third of the clients as malicious, adhering to the maximum Byzantine fault tolerance typically allowed in a distributed system. As depicted in Fig. \ref{fig:byzantine}, the performance of our system remains unaffected by these Byzantine clients, demonstrating its robustness in such scenarios.

\begin{figure}[t]
    \centering
    \includegraphics[width=\linewidth]{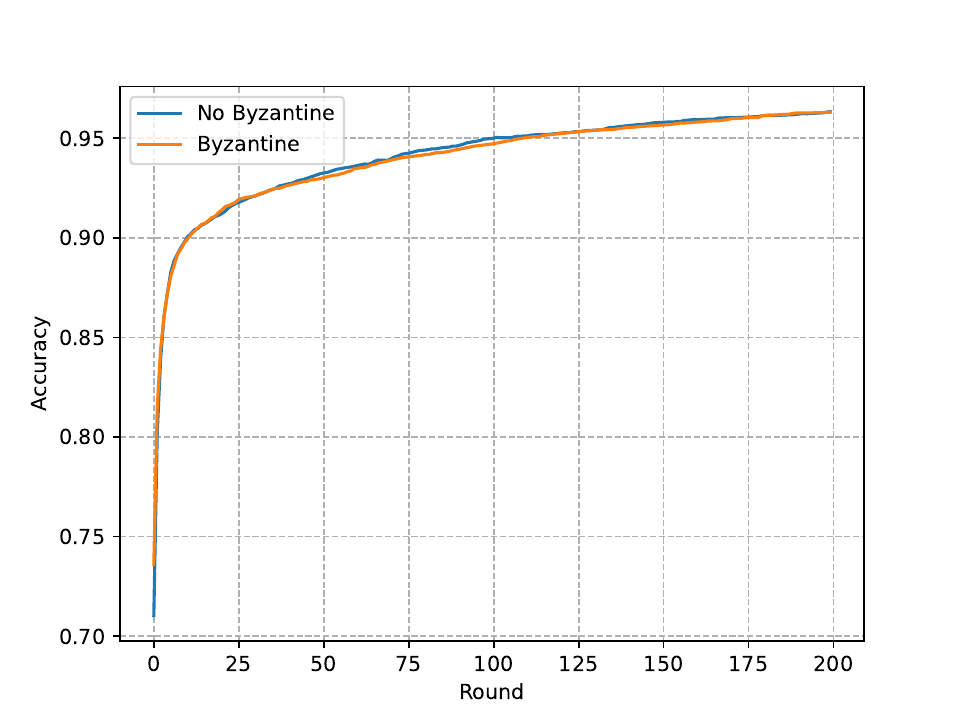}
    \caption{Test accuracy of ZKPoT under honest client settings for MNIST}
    \label{fig:byzantine mnist}
\end{figure}

\begin{table*}[t]
\centering
\caption{Breakdown of time (s) in different phases in ZKPoT under different settings when processing CIFAR10 and MNIST datasets}
\label{breakdown}
\begin{tabularx}{\textwidth}{>{\bfseries}l *{4}{>{\centering\arraybackslash}X}}
\toprule
& \multicolumn{2}{c}{CIFAR10} & \multicolumn{2}{c}{MNIST} \\
\cmidrule(lr){2-3} \cmidrule(lr){4-5}
Stage & (100 nodes) & (200 nodes) & (100 nodes) & (200 nodes) \\
\midrule
Setup          & 204.497 & 204.497 & 193.256 & 193.256 \\
Commitment     & 0.481   & 0.571   & 0.679   & 0.718 \\
Prove          & 195.378 & 198.128 & 168.125 & 172.548 \\
Verify         & 4.855   & 6.759  & 4.790   & 6.293 \\
Block Generation & 0.593 & 0.653   & 0.160   & 0.229 \\
Total Time (s)    & 405.804  & 410.608  & 368.01 & 373.044 \\
\bottomrule
\end{tabularx}
\end{table*}

\begin{figure}[t]
\vspace{-0.2cm}
    \centering
    \includegraphics[width=\linewidth]{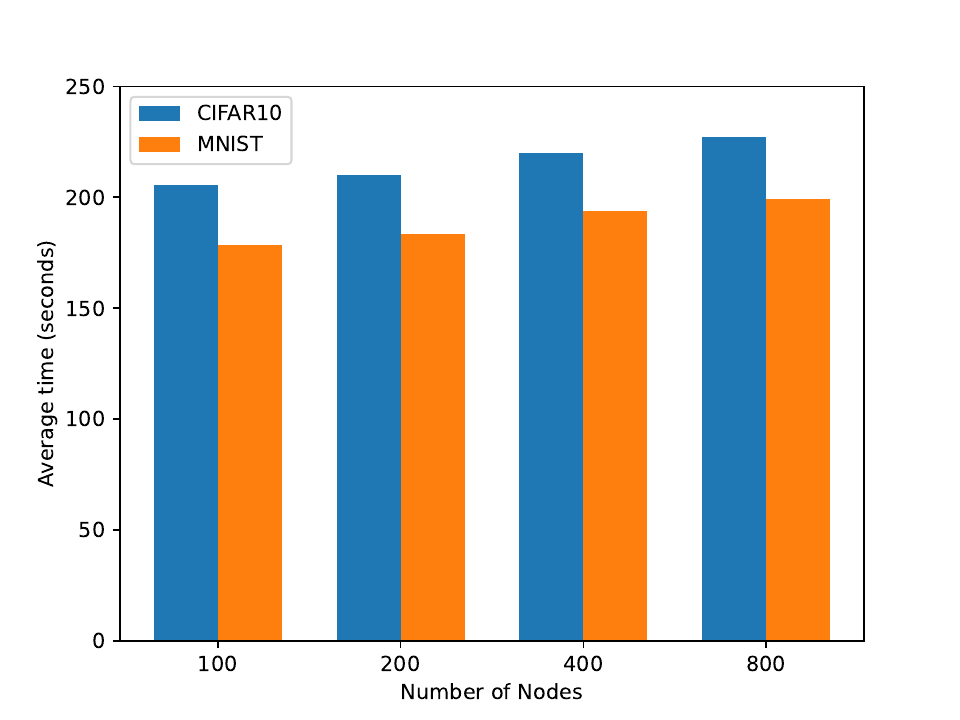}
    \caption{The average time of per verified block generation with different blockchain network size for CIFAR10 and MNIST tasks}
    \label{fig:time}
\end{figure}

\subsection{System Analysis of ZKPoT}
In this section, we assess the computational overhead associated with the key stages of ZKPoT and explore its performance implications as the blockchain network scales.

\textbf{Overhead breakdown:} To quantify the overhead associated with the principal stages of ZKPoT, we simulate our system across a range of blockchain nodes. The number of nodes corresponds to varying numbers of clients enrolled in the FL task in our experiments. We recorded the time spent in the major stages, beginning with the setup phase: 1) setup: the task publisher executes \textit{ZKPoT.Setup()} to generate the proving and verification keys for the model used in the current FL task. This task needs to be executed only once, as the model structure remains unchanged throughout the FL task; 2) commitment: each client participating in the training commits to its local model; 3) prove: clients with trained local models generate a proof $\pi_{acc}$, using public test data $D$ and the corresponding truth label $T$; 4) verify: all nodes verify the transactions in the sorted transaction pool using \textit{ZKPoT.Verify()} until they identify the first valid transaction; 5) block generation: the selected leader creates a new block that records all valid transactions and the new global model. As shown in Table \ref{breakdown}, we deployed ZKPoT in blockchain networks with 100 and 200 nodes for both FL tasks. The recorded values represent the average time (in seconds) for 50 training rounds, excluding the setup phase. The setup phase, which takes the longest time, is performed only once by the task publisher during the first round, and the generated CRS can be reused if the structure of the global model remains unchanged. Additionally, the setup time is influenced solely by the model used for training and remains unaffected by the size of the network. The time required for commitment and proving increases slightly as the number of nodes grows, primarily due to transmission costs. The verification stage takes longer because more votes are needed to verify proof, although it still remains at a very low level. Finally, the block generation stage time increases slightly as more transactions need to be incorporated.

\textbf{Scalability Analysis:} To further assess the scalability of ZKPoT, we ran CIFAR10 and MNIST tasks with varying numbers of blockchain nodes for 50 training rounds and computed the average time required to generate a new block. As shown in Fig. \ref{fig:time}, the average time to generate a valid block increases marginally as the number of blockchain nodes rises from 100 to 800, but the growth remains insignificant. Additionally, the time required is higher for the CIFAR10 task compared to MNIST, due to the utilization of a more sophisticated model, which complicates the proving and verification processes.

\section{Conclusion}\label{sec:conclusion}
In this paper, we proposed ZKPoT with a secure and efficient consensus mechanism for scalable blockchain-secured FL systems. To reduce energy and computation resource consumption, we replaced the traditional mining process with FL training. Specifically, we implemented a zk-SNARK protocol to protect the privacy of local models. The proving and verification processes are optimized by incorporating quantization techniques and having the task publisher actively assist in these stages. We conducted the security analysis for the proposed consensus mechanism and demonstrated that the likelihood of leaking model information during the training and verification process is negligible. The experimental results show that ZKPoT achieves robustness against privacy attacks without degrading the global model performance, and maintains high scalability on blockchain networks of varied sizes.

% use section* for acknowledgment
% \section*{Acknowledgment}

% Can use something like this to put references on a page
% by themselves when using endfloat and the captionsoff option.
\ifCLASSOPTIONcaptionsoff
  \newpage
\fi

% references section

% can use a bibliography generated by BibTeX as a .bbl file
% BibTeX documentation can be easily obtained at:
% http://mirror.ctan.org/biblio/bibtex/contrib/doc/
% The IEEEtran BibTeX style support page is at:
% http://www.michaelshell.org/tex/ieeetran/bibtex/
%\bibliographystyle{IEEEtran}
% argument is your BibTeX string definitions and bibliography database(s)
%\bibliography{IEEEabrv,../bib/paper}
%
% <OR> manually copy in the resultant .bbl file
% set second argument of \begin to the number of references
% (used to reserve space for the reference number labels box)
% \begin{thebibliography}{1}
% % \bibitem{IEEEhowto:kopka}
% % H.~Kopka and P.~W. Daly, \emph{A Guide to \LaTeX}, 3rd~ed.\hskip 1em plus
% %   0.5em minus 0.4em\relax Harlow, England: Addison-Wesley, 1999.
% \end{thebibliography}
\bibliographystyle{IEEEtran}
\bibliography{main}

% Generated by IEEEtran.bst, version: 1.14 (2015/08/26)
\begin{thebibliography}{10}
\providecommand{\url}[1]{#1}
\csname url@samestyle\endcsname
\providecommand{\newblock}{\relax}
\providecommand{\bibinfo}[2]{#2}
\providecommand{\BIBentrySTDinterwordspacing}{\spaceskip=0pt\relax}
\providecommand{\BIBentryALTinterwordstretchfactor}{4}
\providecommand{\BIBentryALTinterwordspacing}{\spaceskip=\fontdimen2\font plus
\BIBentryALTinterwordstretchfactor\fontdimen3\font minus \fontdimen4\font\relax}
\providecommand{\BIBforeignlanguage}[2]{{%
\expandafter\ifx\csname l@#1\endcsname\relax
\typeout{** WARNING: IEEEtran.bst: No hyphenation pattern has been}%
\typeout{** loaded for the language `#1'. Using the pattern for}%
\typeout{** the default language instead.}%
\else
\language=\csname l@#1\endcsname
\fi
#2}}
\providecommand{\BIBdecl}{\relax}
\BIBdecl

\bibitem{al2019privacy}
M.~Al-Rubaie and J.~M. Chang, ``Privacy-preserving machine learning: Threats and solutions,'' \emph{IEEE Security \& Privacy}, vol.~17, no.~2, pp. 49--58, 2019.

\bibitem{fedavg}
B.~McMahan, E.~Moore, D.~Ramage, S.~Hampson, and B.~A. y~Arcas, ``Communication-efficient learning of deep networks from decentralized data,'' in \emph{Artificial intelligence and statistics}.\hskip 1em plus 0.5em minus 0.4em\relax PMLR, 2017, pp. 1273--1282.

\bibitem{acm_survey}
Y.~Qu, M.~P. Uddin, C.~Gan, Y.~Xiang, L.~Gao, and J.~Yearwood, ``Blockchain-enabled federated learning: A survey,'' \emph{ACM Computing Surveys}, vol.~55, no.~4, pp. 1--35, 2022.

\bibitem{rao2024privacy}
B.~Rao, J.~Zhang, D.~Wu, C.~Zhu, X.~Sun, and B.~Chen, ``Privacy inference attack and defense in centralized and federated learning: A comprehensive survey,'' \emph{IEEE Transactions on Artificial Intelligence}, 2024.

\bibitem{nguyen2021federated}
D.~C. Nguyen, M.~Ding, Q.-V. Pham, P.~N. Pathirana, L.~B. Le, A.~Seneviratne, J.~Li, D.~Niyato, and H.~V. Poor, ``Federated learning meets blockchain in edge computing: Opportunities and challenges,'' \emph{IEEE Internet of Things Journal}, vol.~8, no.~16, pp. 12\,806--12\,825, 2021.

\bibitem{feng2021blockchain}
L.~Feng, Y.~Zhao, S.~Guo, X.~Qiu, W.~Li, and P.~Yu, ``Blockchain-based asynchronous federated learning for internet of things,'' \emph{IEEE Transactions on Computers}, vol.~99, no.~1, pp. 1--9, 2021.

\bibitem{jin2023lightweight}
R.~Jin, J.~Hu, G.~Min, and J.~Mills, ``Lightweight blockchain-empowered secure and efficient federated edge learning,'' \emph{IEEE Transactions on Computers}, 2023.

\bibitem{Blockchained_on_device}
H.~Kim, J.~Park, M.~Bennis, and S.-L. Kim, ``Blockchained on-device federated learning,'' \emph{IEEE Communications Letters}, vol.~24, no.~6, pp. 1279--1283, 2019.

\bibitem{biscotti}
M.~Shayan, C.~Fung, C.~J. Yoon, and I.~Beschastnikh, ``Biscotti: A blockchain system for private and secure federated learning,'' \emph{IEEE Transactions on Parallel and Distributed Systems}, vol.~32, no.~7, pp. 1513--1525, 2020.

\bibitem{commitee_consensus}
Y.~Li, C.~Chen, N.~Liu, H.~Huang, Z.~Zheng, and Q.~Yan, ``A blockchain-based decentralized federated learning framework with committee consensus,'' \emph{IEEE Network}, vol.~35, no.~1, pp. 234--241, 2020.

\bibitem{feng2021bafl}
L.~Feng, Y.~Zhao, S.~Guo, X.~Qiu, W.~Li, and P.~Yu, ``Bafl: A blockchain-based asynchronous federated learning framework,'' \emph{IEEE Transactions on Computers}, vol.~71, no.~5, pp. 1092--1103, 2021.

\bibitem{yang2024blockchain}
R.~Yang, T.~Zhao, F.~R. Yu, M.~Li, D.~Zhang, and X.~Zhao, ``Blockchain-based federated learning with enhanced privacy and security using homomorphic encryption and reputation,'' \emph{IEEE Internet of Things Journal}, 2024.

\bibitem{xu2022spdl}
M.~Xu, Z.~Zou, Y.~Cheng, Q.~Hu, D.~Yu, and X.~Cheng, ``Spdl: A blockchain-enabled secure and privacy-preserving decentralized learning system,'' \emph{IEEE Transactions on Computers}, vol.~72, no.~2, pp. 548--558, 2022.

\bibitem{cognitive_com}
Y.~Qu, S.~R. Pokhrel, S.~Garg, L.~Gao, and Y.~Xiang, ``A blockchained federated learning framework for cognitive computing in industry 4.0 networks,'' \emph{IEEE Transactions on Industrial Informatics}, vol.~17, no.~4, pp. 2964--2973, 2020.

\bibitem{podl}
C.~Chenli, B.~Li, Y.~Shi, and T.~Jung, ``Energy-recycling blockchain with proof-of-deep-learning,'' in \emph{2019 IEEE International Conference on Blockchain and Cryptocurrency (ICBC)}.\hskip 1em plus 0.5em minus 0.4em\relax IEEE, 2019, pp. 19--23.

\bibitem{pofl}
X.~Qu, S.~Wang, Q.~Hu, and X.~Cheng, ``Proof of federated learning: A novel energy-recycling consensus algorithm,'' \emph{IEEE Transactions on Parallel and Distributed Systems}, vol.~32, no.~8, pp. 2074--2085, 2021.

\bibitem{DLG}
L.~Zhu, Z.~Liu, and S.~Han, ``Deep leakage from gradients,'' \emph{Advances in neural information processing systems}, vol.~32, 2019.

\bibitem{PoQ}
Y.~Lu, X.~Huang, Y.~Dai, S.~Maharjan, and Y.~Zhang, ``Blockchain and federated learning for privacy-preserved data sharing in industrial iot,'' \emph{IEEE Transactions on Industrial Informatics}, vol.~16, no.~6, pp. 4177--4186, 2019.

\bibitem{pf-pofl}
Y.~Wang, H.~Peng, Z.~Su, T.~H. Luan, A.~Benslimane, and Y.~Wu, ``A platform-free proof of federated learning consensus mechanism for sustainable blockchains,'' \emph{IEEE Journal on Selected Areas in Communications}, vol.~40, no.~12, pp. 3305--3324, 2022.

\bibitem{wei2020federated}
K.~Wei, J.~Li, M.~Ding, C.~Ma, H.~H. Yang, F.~Farokhi, S.~Jin, T.~Q. Quek, and H.~V. Poor, ``Federated learning with differential privacy: Algorithms and performance analysis,'' \emph{IEEE transactions on information forensics and security}, vol.~15, pp. 3454--3469, 2020.

\bibitem{DMIA}
S.~Truex, L.~Liu, M.~E. Gursoy, L.~Yu, and W.~Wei, ``Demystifying membership inference attacks in machine learning as a service,'' \emph{IEEE transactions on services computing}, vol.~14, no.~6, pp. 2073--2089, 2019.

\bibitem{gupta2022recovering}
S.~Gupta, Y.~Huang, Z.~Zhong, T.~Gao, K.~Li, and D.~Chen, ``Recovering private text in federated learning of language models,'' \emph{Advances in neural information processing systems}, vol.~35, pp. 8130--8143, 2022.

\bibitem{zkcnn}
T.~Liu, X.~Xie, and Y.~Zhang, ``Zkcnn: Zero knowledge proofs for convolutional neural network predictions and accuracy,'' in \emph{Proceedings of the 2021 ACM SIGSAC Conference on Computer and Communications Security}, 2021, pp. 2968--2985.

\bibitem{fan2024validcnn}
Y.~Fan, K.~Ma, L.~Zhang, X.~Lei, G.~Xu, and G.~Tan, ``Validcnn: A large-scale cnn predictive integrity verification scheme based on zk-snark,'' \emph{IEEE Transactions on Dependable and Secure Computing}, 2024.

\bibitem{dpos_iov}
Y.~Lu, X.~Huang, K.~Zhang, S.~Maharjan, and Y.~Zhang, ``Blockchain empowered asynchronous federated learning for secure data sharing in internet of vehicles,'' \emph{IEEE Transactions on Vehicular Technology}, vol.~69, no.~4, pp. 4298--4311, 2020.

\bibitem{pos_iot}
Y.~Zhao, J.~Zhao, L.~Jiang, R.~Tan, D.~Niyato, Z.~Li, L.~Lyu, and Y.~Liu, ``Privacy-preserving blockchain-based federated learning for iot devices,'' \emph{IEEE Internet of Things Journal}, vol.~8, no.~3, pp. 1817--1829, 2020.

\bibitem{IDLG}
B.~Zhao, K.~R. Mopuri, and H.~Bilen, ``idlg: Improved deep leakage from gradients,'' \emph{arXiv preprint arXiv:2001.02610}, 2020.

\bibitem{MIA}
M.~Bertran, S.~Tang, A.~Roth, M.~Kearns, J.~H. Morgenstern, and S.~Z. Wu, ``Scalable membership inference attacks via quantile regression,'' \emph{Advances in Neural Information Processing Systems}, vol.~36, 2024.

\bibitem{saftynet}
Z.~Ghodsi, T.~Gu, and S.~Garg, ``Safetynets: Verifiable execution of deep neural networks on an untrusted cloud,'' \emph{Advances in Neural Information Processing Systems}, vol.~30, 2017.

\bibitem{vcnn}
S.~Lee, H.~Ko, J.~Kim, and H.~Oh, ``vcnn: Verifiable convolutional neural network based on zk-snarks,'' \emph{IEEE Transactions on Dependable and Secure Computing}, 2024.

\bibitem{zen}
B.~Feng, L.~Qin, Z.~Zhang, Y.~Ding, and S.~Chu, ``Zen: An optimizing compiler for verifiable, zero-knowledge neural network inferences,'' \emph{Cryptology ePrint Archive}, 2021.

\bibitem{zkfl}
Z.~Wang, N.~Dong, J.~Sun, and W.~Knottenbelt, ``zkfl: Zero-knowledge proof-based gradient aggregation for federated learning,'' \emph{arXiv preprint arXiv:2310.02554}, 2023.

\bibitem{zkfdl}
M.~Ahmadi and R.~Nourmohammadi, ``zkfdl: An efficient and privacy-preserving decentralized federated learning with zero knowledge proof,'' in \emph{2024 IEEE 3rd International Conference on AI in Cybersecurity (ICAIC)}.\hskip 1em plus 0.5em minus 0.4em\relax IEEE, 2024, pp. 1--10.

\bibitem{zkp-fl}
Z.~Xing, Z.~Zhang, M.~Li, J.~Liu, L.~Zhu, G.~Russello, and M.~R. Asghar, ``Zero-knowledge proof-based practical federated learning on blockchain,'' \emph{arXiv preprint arXiv:2304.05590}, 2023.

\bibitem{zpol}
H.~Zhang, J.~Wu, X.~Lin, A.~K. Bashir, and Y.~D. Al-Otaibi, ``Integrating blockchain and deep learning into extremely resource-constrained iot: an energy-saving zero-knowledge pol approach,'' \emph{IEEE Internet of Things Journal}, 2023.

\bibitem{ipfs}
J.~Benet, ``Ipfs-content addressed, versioned, p2p file system,'' \emph{arXiv preprint arXiv:1407.3561}, 2014.

\bibitem{jebreel2022enhanced}
N.~M. Jebreel, J.~Domingo-Ferrer, A.~Blanco-Justicia, and D.~S{\'a}nchez, ``Enhanced security and privacy via fragmented federated learning,'' \emph{IEEE Transactions on Neural Networks and Learning Systems}, 2022.

\bibitem{zhu2023blockchain}
J.~Zhu, J.~Cao, D.~Saxena, S.~Jiang, and H.~Ferradi, ``Blockchain-empowered federated learning: Challenges, solutions, and future directions,'' \emph{ACM Computing Surveys}, vol.~55, no.~11, pp. 1--31, 2023.

\bibitem{groth16}
J.~Groth, ``On the size of pairing-based non-interactive arguments,'' in \emph{Advances in Cryptology--EUROCRYPT 2016: 35th Annual International Conference on the Theory and Applications of Cryptographic Techniques, Vienna, Austria, May 8-12, 2016, Proceedings, Part II 35}.\hskip 1em plus 0.5em minus 0.4em\relax Springer, 2016, pp. 305--326.

\bibitem{opacus}
A.~Yousefpour, I.~Shilov, A.~Sablayrolles, D.~Testuggine, K.~Prasad, M.~Malek, J.~Nguyen, S.~Ghosh, A.~Bharadwaj, J.~Zhao \emph{et~al.}, ``Opacus: User-friendly differential privacy library in pytorch,'' \emph{arXiv preprint arXiv:2109.12298}, 2021.

\bibitem{gaussianmip}
T.~Leemann, M.~Pawelczyk, and G.~Kasneci, ``Gaussian membership inference privacy,'' \emph{Advances in Neural Information Processing Systems}, vol.~36, 2024.

\bibitem{geiping2020inverting}
J.~Geiping, H.~Bauermeister, H.~Dr{\"o}ge, and M.~Moeller, ``Inverting gradients-how easy is it to break privacy in federated learning?'' \emph{Advances in neural information processing systems}, vol.~33, pp. 16\,937--16\,947, 2020.

\bibitem{i2f}
H.~Zhang, J.~Hong, Y.~Deng, M.~Mahdavi, and J.~Zhou, ``Understanding deep gradient leakage via inversion influence functions,'' \emph{Advances in Neural Information Processing Systems}, vol.~36, 2024.

\end{thebibliography}

% that's all folks
\end{document}